\documentclass[11pt,twoside]{article}
\usepackage{graphicx,epsfig,natbib,epstopdf}
\usepackage{CS18}

\usepackage{amssymb,amsmath}

\begin{document}
%
%
%
\title{Asteroseismology\index{asteroseismology} of Cool Stars}
%
%
\author{Daniel Huber$^{1,2,3}$}
\affil{$^1$NASA Ames Research Center, Moffett Field, CA 94035, USA}
\affil{$^2$SETI Institute, 189 Bernardo Avenue, Mountain View, CA 94043, USA}
\affil{$^3$Sydney Institute for Astronomy (SIfA), School of Physics, University of Sydney, 
NSW 2006, Australia; dhuber@physics.usyd.edu.au}

\begin{abstract}
%
%
The measurement of oscillations excited by surface convection is a powerful method to study 
the structure and evolution of cool stars. CoRoT and \textit{Kepler} have initiated a 
revolution in asteroseismology by detecting oscillations in thousands of stars from the 
main sequence to the red-giant branch, including a large number of exoplanet host stars. 
In this contribution I will review recent asteroseismic results, focusing in particular on 
the internal rotation of red giant stars and 
the impact of asteroseismology on the characterization of exoplanets.
\end{abstract}
%
%
%
%
%

\section{Introduction\index{stellar oscillations}}

\begin{figure}[t!]
\begin{center}
\resizebox{12cm}{!}{\includegraphics{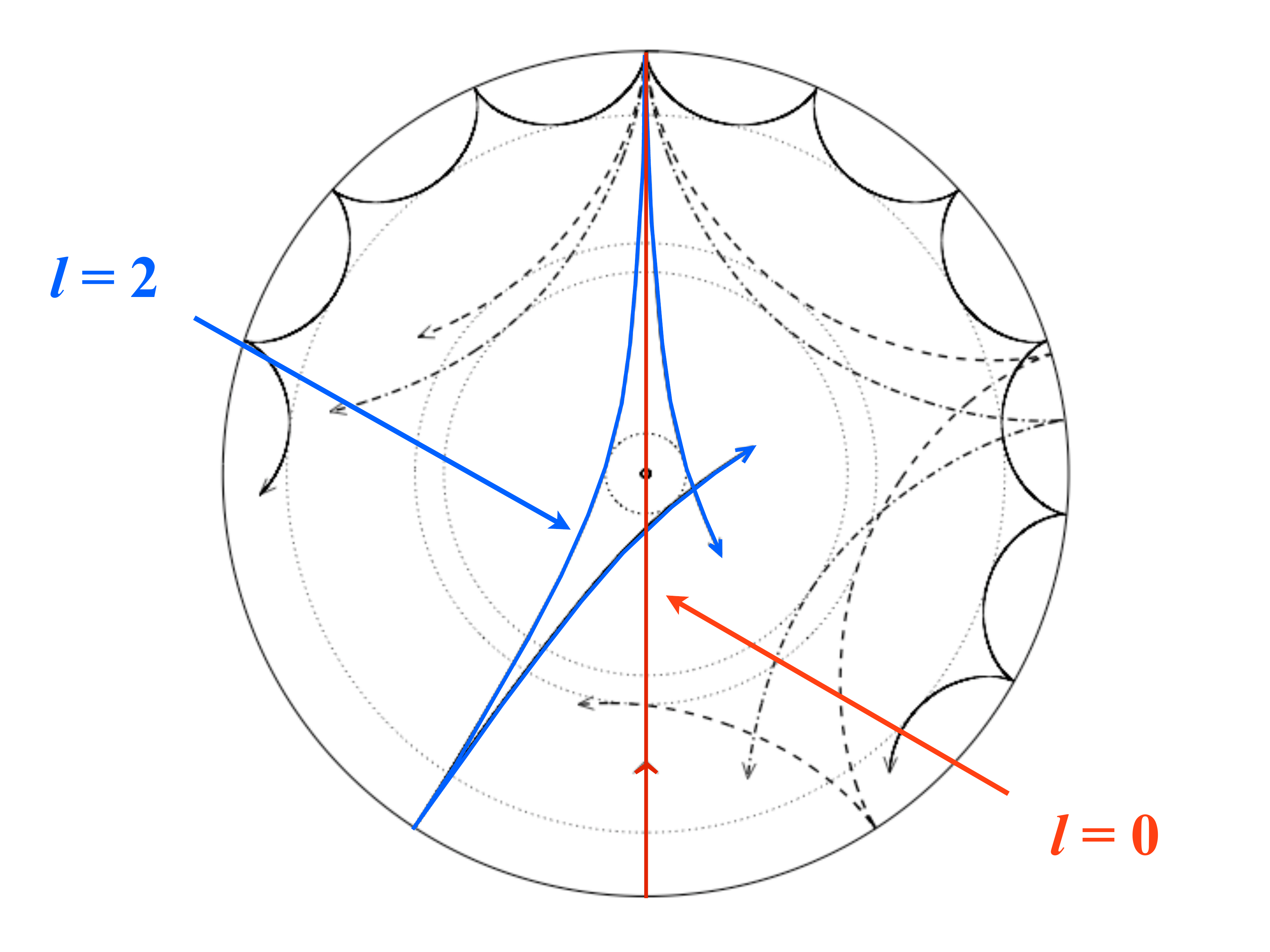}}
\caption{Schematic cross-section of a star illustrating the paths 
of modes with different spherical degrees $l$. Adapted from \citet{CD03}.}
\label{seismo:fig1}
\end{center}
\end{figure}

Stellar oscillations in cool stars are excited by turbulent convection in the outer 
layers \citep[e.g.][]{houdek99}. Oscillation modes 
can be described by spherical harmonics with spherical degree $l$ 
(the total number of node lines on the surface), azimuthal order 
$|m|$ (the number of node lines that cross the equator), and radial order 
$n$ (the number of nodes from the surface to the center of the star).
Radial pulsations are therefore expressed as $l=0$, while $l>0$ 
are non-radial pulsations. Modes with higher spherical degrees penetrate to shallower depths 
within the star (Figure \ref{seismo:fig1}).

Oscillation modes can furthermore be separated into 
pressure modes (p modes) and gravity modes (g modes). 
Pressure modes are acoustic waves propagating through the 
compression and 
decompression of gas, and pressure gradient acts as the restoring force. Gravity 
modes are pulsations due to the interplay of 
buoyancy and gravity, and buoyancy acts the 
restoring force. Gravity modes are damped where 
convection is unstable, and are therefore usually confined to the deep interior for cool stars. 
Pressure modes propagate in radiative zones, and hence 
are more easily excited to observational amplitudes on the surface.

Oscillation frequencies $\nu_{n,l}$ of high $n$ and 
low $l$ can be described by the asymptotic theory of stellar 
oscillations \citep{vandakurov68,tassoul80,gough86}, which observationally 
can be approximated as follows:

\begin{equation}
\nu_{n,l} \approx \Delta\nu(n + \frac{1}{2}l + \epsilon) - \delta\nu_{0l} \: .
\label{equ:asymt}
\end{equation}

\noindent
Here, the large frequency separation
$\Delta \nu$ is the separation of modes of the same spherical degree $l$ 
and consecutive radial order $n$, 
while modes with different degree $l$ and same 
order $n$ are separated by the small frequency separations $\delta\nu_{0l}$. 
In the asymptotic theory, $\Delta \nu$ can be shown to be equal to the inverse of twice 
the sound travel time through the stellar diameter \citep{ulrich86,CD03}:

\begin{equation}
\Delta\nu = \left(2 \int^{R}_{0} \frac{dr}{c} \right)^{-1} \: ,
\label{seismo:dnu_phys}
\end{equation}

\noindent
where $c$ is the sound speed. Assuming adiabacity and an ideal gas 
$c \propto \sqrt{T/\mu}$ and $T\propto \mu M / R$, where $\mu$ is the mean 
molecular weight. Hence, Equation (\ref{seismo:dnu_phys}) 
can be expressed as \citep{KB95}:

\begin{equation}
\Delta\nu \propto \left(\frac{M}{R^3}\right)^{1/2} \: .
\end{equation}

\noindent
The large frequency separation is therefore a measure of the mean stellar density. 
The small frequency separations can be written as \citep{CD03}:

\begin{equation}
\delta \nu_{nl} = -(4l+6) \frac{\Delta\nu}{4\pi^2\nu_{nl}} \int^{R}_{0} \frac{dc}{dr}\frac{dr}{r} \: .
\label{seismo:smallsp}
\end{equation}
 
\noindent
The integral shows that $\delta \nu_{nl}$ is sensitive to sound-speed gradient in 
the stellar interior, which depends on the chemical composition profile. Hence, 
$\delta \nu_{nl}$ is sensitive to changes in the chemical composition during stellar evolution 
and thus stellar age. Equation (\ref{seismo:smallsp}) can be qualitatively understood by the 
fact that modes of different $l$ travel to different depths within the star, 
and hence their frequency differences provide information about the radial 
structure. 

An additional asteroseismic observable is the frequency of maximum power, $\nu_{\rm max}$, 
which is related to the driving and 
damping of the modes. The frequency of maximum power has been suggested to scale with the 
acoustic cut-off frequency \citep{brown91}, which is the maximum frequency below which an acoustic 
mode can be reflected \citep{CD03}:

\begin{equation}
\nu_{\rm ac} = \frac{c}{2 H_{\rm p}} \: .
\end{equation}

\noindent
Here, $H_{\rm p}$ is the pressure scale height. For an isothermal atmosphere 
$H_{\rm p} = \frac{P R^2}{G M \rho}$, which combined with the ideal gas equation yields:

\begin{equation}
\nu_{\rm max} \propto \nu_{\rm ac} \propto \frac{M}{R^2 \sqrt{T_{\rm eff}}} \: .
\label{equ:numax}
\end{equation}

\noindent
Hence, $\nu_{\rm max}$ is mainly a measure of the surface gravity of a star.

Measurements of $\nu_{\rm max}$, $\Delta \nu$, $\delta \nu_{nl}$, or individual frequencies 
provide a powerful way to determine the stellar structure and fundamental stellar properties 
(such as radius and mass), 
which are otherwise difficult to measure for field stars. Validations of 
asteroseismic results with classical methods such as interferometry and 
eclipsing binary systems have 
shown good agreement within the typically quoted uncertainties 
\citep[see, e.g.,][for recent reviews]{belkacem12,huber14b}.

\section{The Space Photometry Revolution of Asteroseismology\index{CoRoT}\index{Kepler}}

\begin{figure}
\begin{center}
\resizebox{\hsize}{!}{\includegraphics{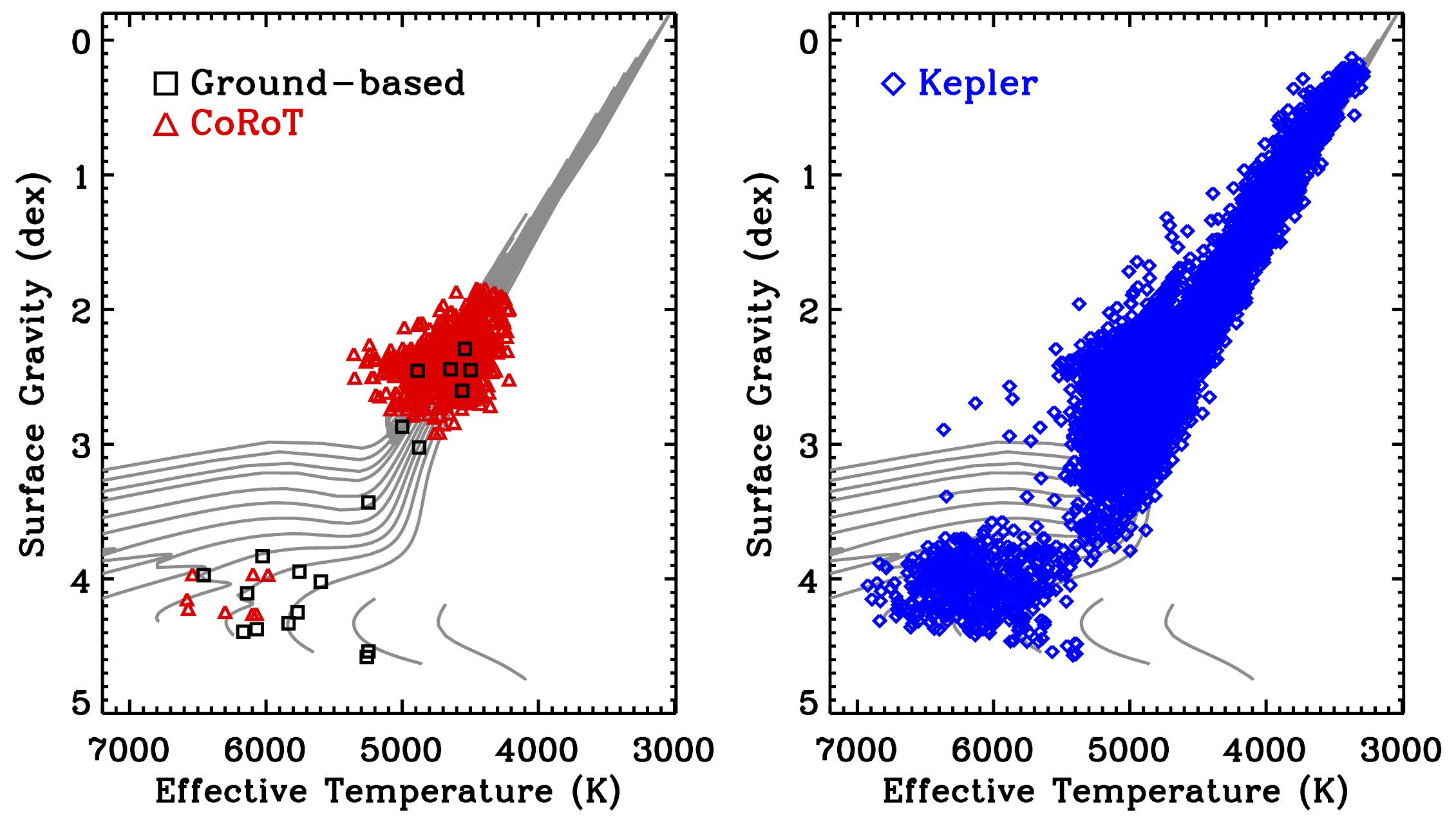}}
\caption{Stars with detected convection-driven oscillations in a $\log (g)$--$T_{\rm{eff}}$\ 
diagram. Left panel: Detections from ground- and space-based observations prior to 2009 
(black squares) and by CoRoT (red triangles). Right panel: Detections by \textit{Kepler} 
taken from \citet{huber14}.}
\label{seismo:fig2}
\end{center}
\end{figure}

Early efforts to detect stellar oscillations used ground-based 
radial-velocity observations. The first confirmed detection 
dates back to \citet{brown91} in Procyon, followed by the first 
detection of regularly spaced frequencies in $\eta$\,Boo by \citet{kjeldsen95}. 
The greatly improved sensitivity of Doppler velocities for detecting exoplanets 
enabled the detection of oscillations in several 
nearby main sequence and subgiant stars such as $\beta$\,Hyi \citep{bedding01,carrier01}, 
$\alpha$\,Cen\,A \citep{bouchy01,butler04} and B \citep{carrier03,kjeldsen05} as well as 
red giant stars such as $\xi$\,Hya \citep{frandsen02} and $\epsilon$\,Oph \citep{deridder06}. 
At the same time, first space-based photometric observations led to detections with 
the Canadian space telescope MOST \citep[Microvariability and Oscillations in Stars,][]{walker03,matthews07} 
in red giants \citep{barban07,kallinger08b} and Procyon \citep{guenther08,huber11}. 
Space-based 
observations were also performed using the startracker of the WIRE 
(Wide-Field Infrared Explorer) satellite \citep{schou01,retter03,bruntt05,stello08}, the 
SMEI (Solar Mass Ejection Imager) experiment \citep{tarrant07} and the 
Hubble Space Telescope \citep{edmonds96,gilliland08,stello09b,gilliland11}.
In total, observational efforts prior to 2009 yielded detections in $\sim 20$ stars 
(see Figure \ref{seismo:fig2}). 

A major breakthrough (and commonly referred to as the beginning of the space photometry 
revolution of asteroseismology) was achieved by the French-led 
CoRoT (Convection Rotation and Planetary Transits) satellite, 
which detected oscillations in a number 
of main sequence stars \citep[e.g.][]{michel08} and several thousands 
red giant stars 
\citep[e.g.][]{hekker09}. Importantly, CoRoT showed that red giants 
oscillate in non-radial modes \citep{deridder09}, which opened the door for detailed 
studies of the interior structure of red giants (see Section \ref{sec:redgiants}).

The \textit{Kepler} space telescope, launched in 2009, completed the 
revolution of cool-star asteroseismology by covering the low-mass H-R 
diagram with detections, including dwarfs cooler than the Sun \citep{chaplin11a} and over 
ten thousand red giants \citep{hekker11c,stello13}. The larger number of red giants with 
detected oscillations is due to a combination of two effects: First, oscillation 
amplitudes increase with luminosity \citep{KB95}, making a detection easier 
at a given apparent magnitude. Second, the majority of \textit{Kepler} targets were 
observed with 30-minute sampling, setting a limit of
$\log g \lesssim 3.5$ since less evolved stars oscillate above the 
Nyquist frequency.

\section{Probing the Cores of Red Giants\index{red giants}\index{rotation}\index{angular momentum}}
\label{sec:redgiants}

For evolved stars the p-mode and g-mode cavity can overlap, giving rise to so-called 
``mixed modes'' \citep{dziembowski01}. Mixed modes contain contributions from g modes 
confined to the core, but 
unlike pure g modes can have low enough mode inertias to be observed at the surface. 
As opposed to p modes, g modes are theoretically predicted to be equally spaced in 
period. The coupling of p modes with g modes causes mixed modes to be shifted from 
their original frequency spacing \citep{aizenman77}, yielding multiple frequencies per 
radial order which are expected to be approximately equally spaced in period. 

The detection of mixed $l=1$ modes in red giants by \textit{Kepler} \citep{bedding10} 
led to several important discoveries in our understanding of the internal composition 
and rotation of giants. Following the observational confirmation of equal period spacings 
\citep{beck11}, \citet{bedding11} demonstrated that giants ascending the RGB and 
He-core burning red giants can be separated based on their mixed-mode 
period spacing \citep[see also][]{mosser11b}. Shortly after, \citet{beck12} showed 
that mixed modes are split into multiplets by rotation, and that frequency splittings 
for g-dominated mixed modes are substantially higher than for p-dominated mixed modes due 
radial differential rotation. 

\begin{figure}
\begin{center}
\resizebox{11cm}{!}{\includegraphics{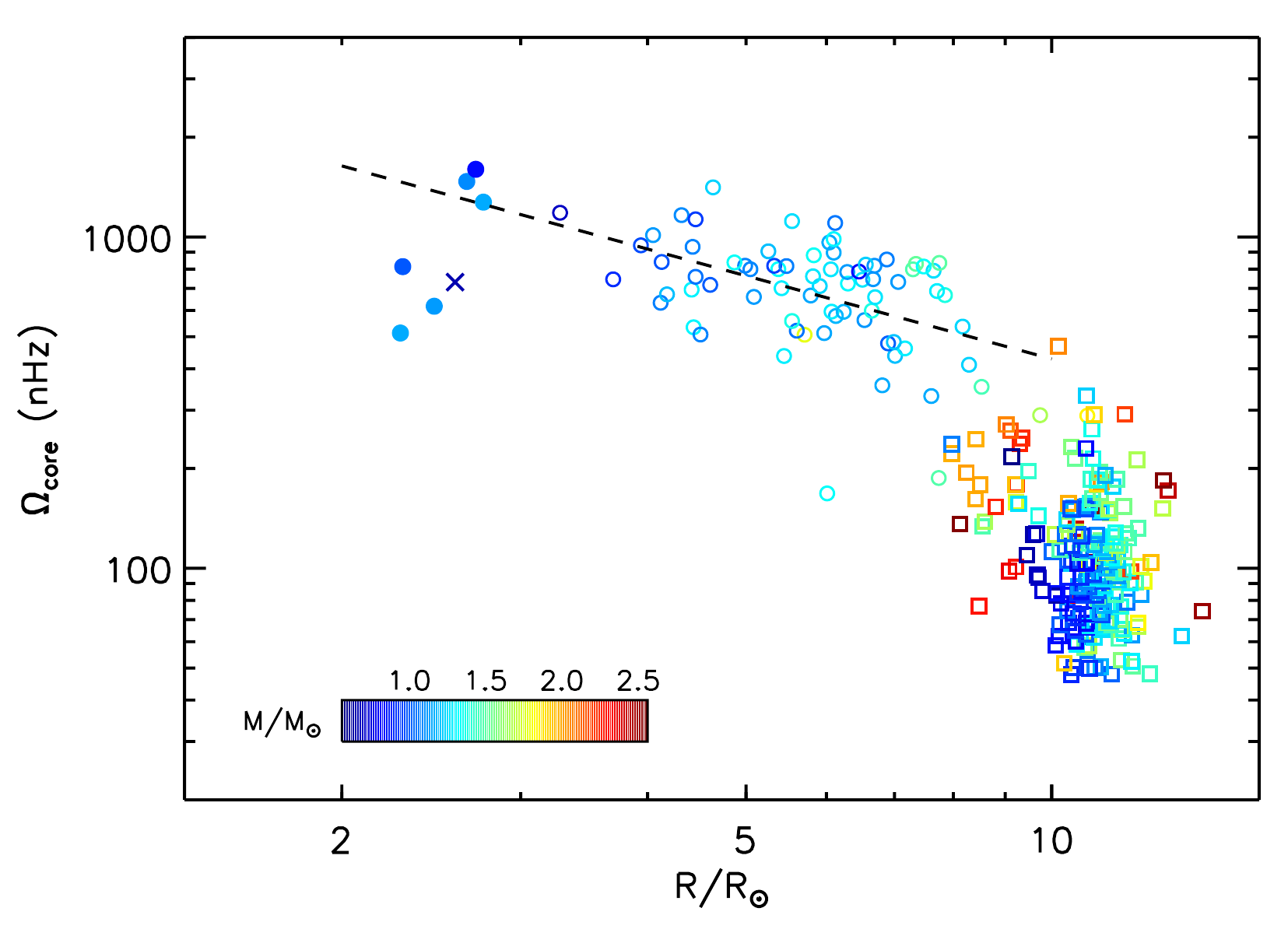}}
\caption{Core rotation frequency versus stellar radius for a sample of subgiants and 
red giants observed by \textit{Kepler}. Circles mark RGB stars, squares He-core burning stars, and 
colors denote the stellar mass determined from asteroseismology. From \citet{deheuvels14}.}
\label{seismo:fig3}
\end{center}
\end{figure}

Core rotation rates were subsequently measured for hundreds of red giants, 
allowing an unprecedented view in the internal rotation evolution of evolved 
stars. 
Figure \ref{seismo:fig3} shows seismically measured core rotation rates as a function of 
stellar radius for a sample of subgiants \citep{deheuvels14} and red giants \citep{mosser13}. 
The data show 
that the cores spin up as stars evolve towards the RGB, followed by a gradual spin-down 
as stars evolve towards the He-core burning main sequence. 
Predicted core rotation rates in models are up to factors of 10--100 larger than 
observed \citep{marques13,cantiello14}, pointing to a yet unidentified mechanism responsible 
for transporting angular momentum from the core to the envelope. The exquisite 
observational data promise significant advances in 
our theoretical understanding of the different roles of 
angular momentum transport mechanisms in stars \citep[e.g.][]{fuller14}.

\section{Asteroseismology of Exoplanet Host Stars\index{exoplanets}}

\subsection{Characterization of Exoplanets}

\begin{figure}
\begin{center}
\resizebox{\hsize}{!}{\includegraphics{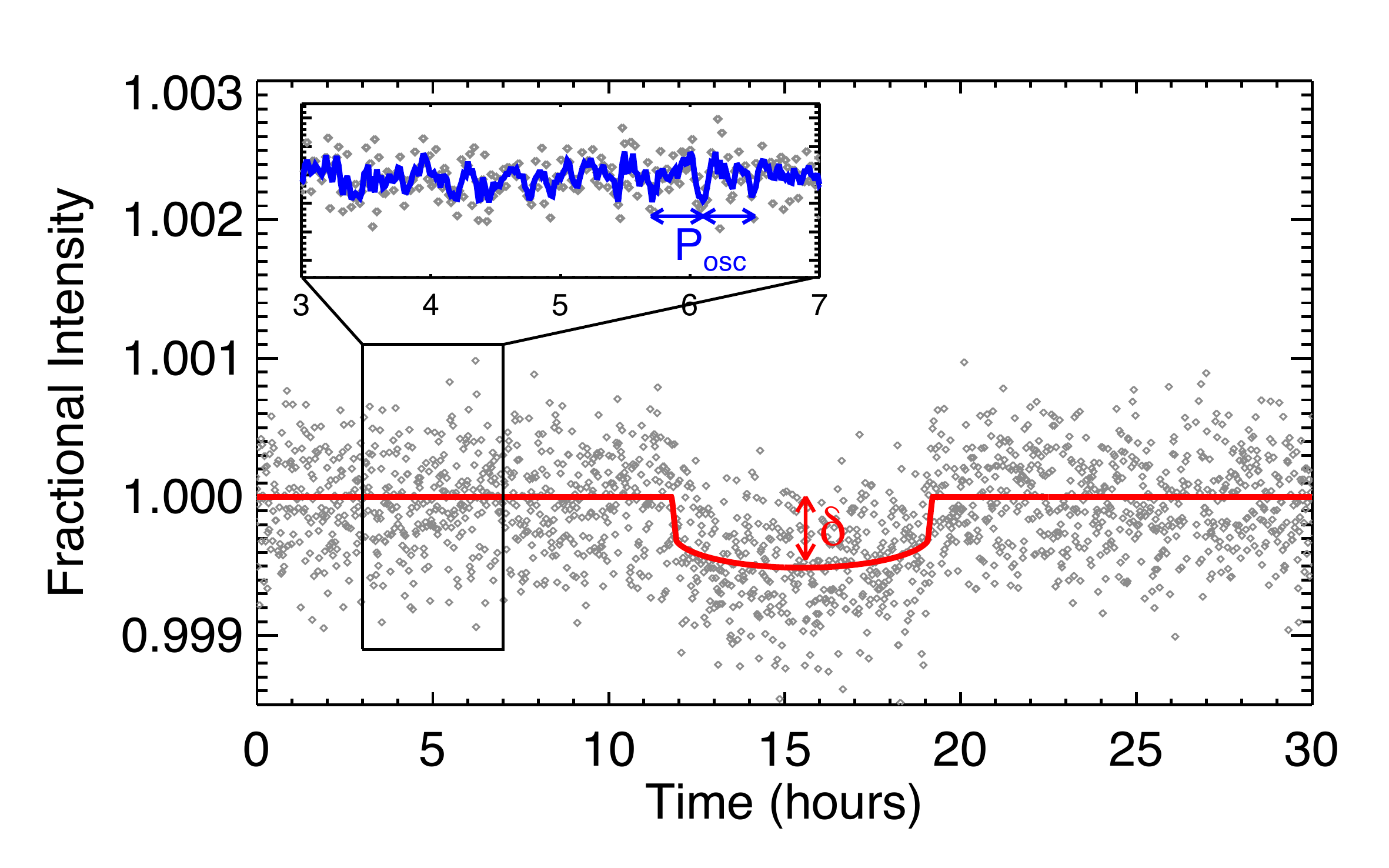}}
\caption{\textit{Kepler} short-cadence light curve centered on a transit of 
Kepler-36c \citep{carter12}. 
The thick solid line shows the best-fitting transit model. 
The inset illustrates the oscillations of the host star.
The transit depth $\delta$ yields the size of the planet relative to the size of the 
star, and the oscillation periods $P_{\rm osc}$ independently measure the size of the star.}
\label{seismo:fig4}
\end{center}
\end{figure}

Transit and radial velocity surveys measure exoplanet properties relative to stellar 
properties, hence requiring a precise characterizations of host stars. 
Asteroseismology is a powerful tool to provide such characterizations in a systematic 
fashion.
In particular, for observations with sufficiently high precision such as provided 
by \textit{Kepler},
both planetary transits and stellar oscillations can be measured using the same data 
(Figure \ref{seismo:fig4}).

\begin{figure}
\begin{center}
\resizebox{12cm}{!}{\includegraphics{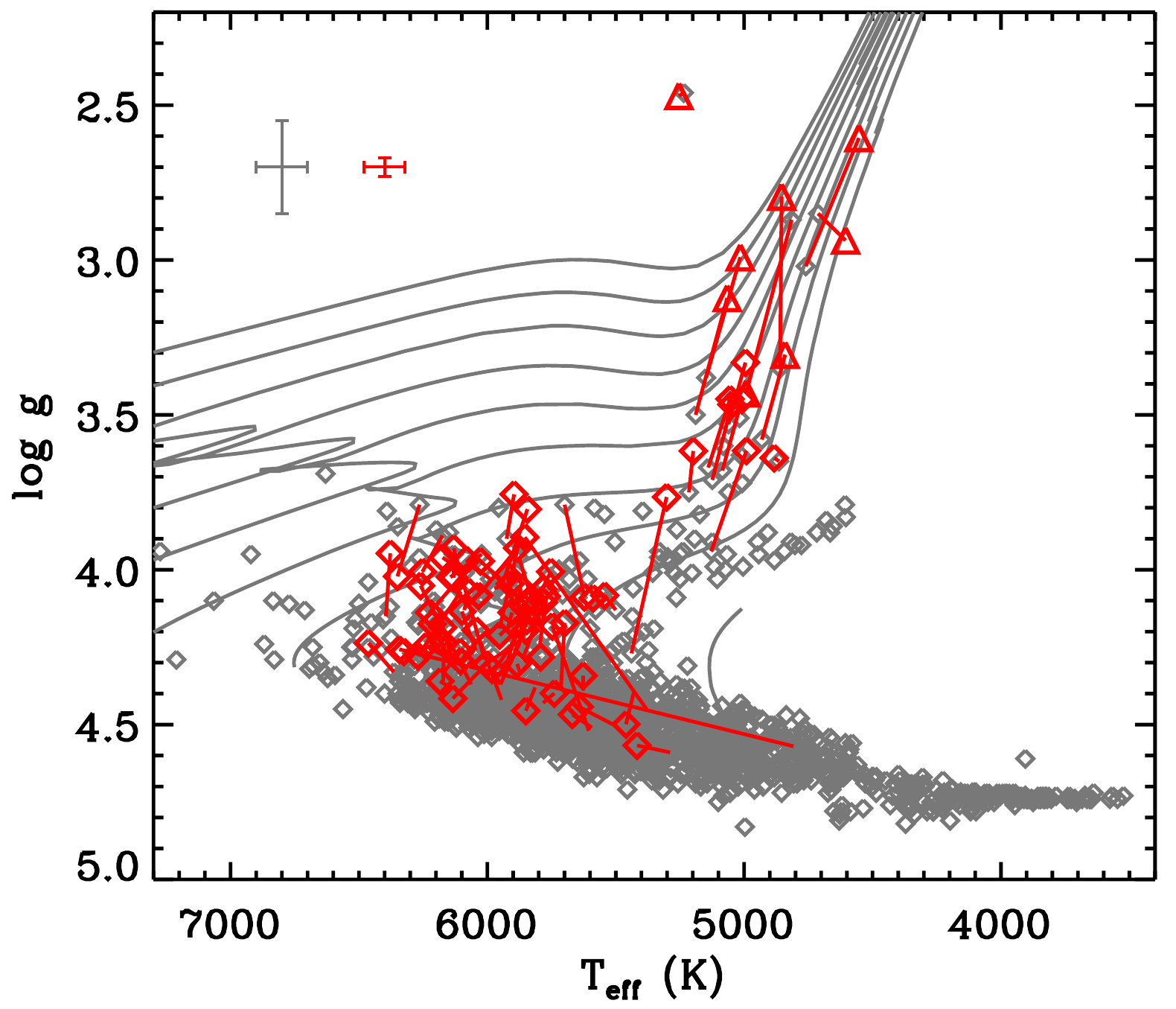}}
\caption{Surface gravity versus effective temperature for planet-candidate hosts 
in \citet{batalha12} (grey diamonds). Solar metallicity evolutionary tracks 
are shown as grey lines. 
Thick red symbols show positions of 77 host stars with asteroseismic 
detections using long-cadence (triangles) and short-cadence (diamonds) data, 
respectively. Red lines 
connect the revised positions to the values in \citet{batalha12}. 
Typical error bars are 
shown in the top left side of the plot. From \citet{huber13}.}
\label{seismo:fig5}
\end{center}
\end{figure}

To date, nearly 80 \textit{Kepler} host stars have been uniformly characterized using 
measurements of $\nu_{\rm max}$ and $\Delta \nu$, yielding precise  ($\sim$ 3\%) radii for 
over 100 exoplanet candidates \citep[Figure \ref{seismo:fig5},][]{huber13}. For a smaller sample 
detailed frequency modeling has yielded even more precise radii, masses and
ages with typical uncertainties of 1\%, 2\% and 10\%, respectively 
\citep[e.g.][]{cd10,gilliland13}. 
Efforts to derive precise and accurate seismic ages for more than 30 \textit{Kepler} host 
stars are currently under way (Davis et al, in prep; Silva Aguirre et al., in prep).

Comparisons with high-resolution spectroscopy for \textit{Kepler} hosts showed that previous 
estimates of $\log g$ for evolved host stars were significantly overestimated, 
while characterizations for dwarfs generally show good agreement 
(Figure \ref{seismo:fig5}). The majority of 
\textit{Kepler} host stars, however, still rely on photometric classifications from the 
Kepler Input Catalog \citep{brown11}, which have been suggested to suffer from 
biases for dwarfs \citep{verner11,everett13,bastien14}. Asteroseismology 
will continue to play a key role to accurately quantify such biases 
through the calibration of less direct methods which are applicable to larger samples 
such as photometric flicker \citep{bastien13} or spectroscopy.

\subsection{Probing the Architecture of Exoplanet Systems\index{exoplanet obliquities}\index{Kepler-56}}

In addition to rotation periods, rotational splittings 
can be used to determine the line-of-sight inclination of the stellar 
rotation axis by measuring the relative heights of the $l(l+1)$ modes in each 
multiplet \citep{gizon03}. For dipole ($l=1$) modes, a doublet implies a high 
inclination (rotation axis perpendicular to the line of sight), while a triplet 
implies an intermediate inclination. Observing a single peak does not constrain 
the inclination since the star can either have a low inclination 
(pole-on) or show no splittings because of slow rotation.

\begin{figure}[ht!]
\begin{center}
\resizebox{12cm}{!}{\includegraphics{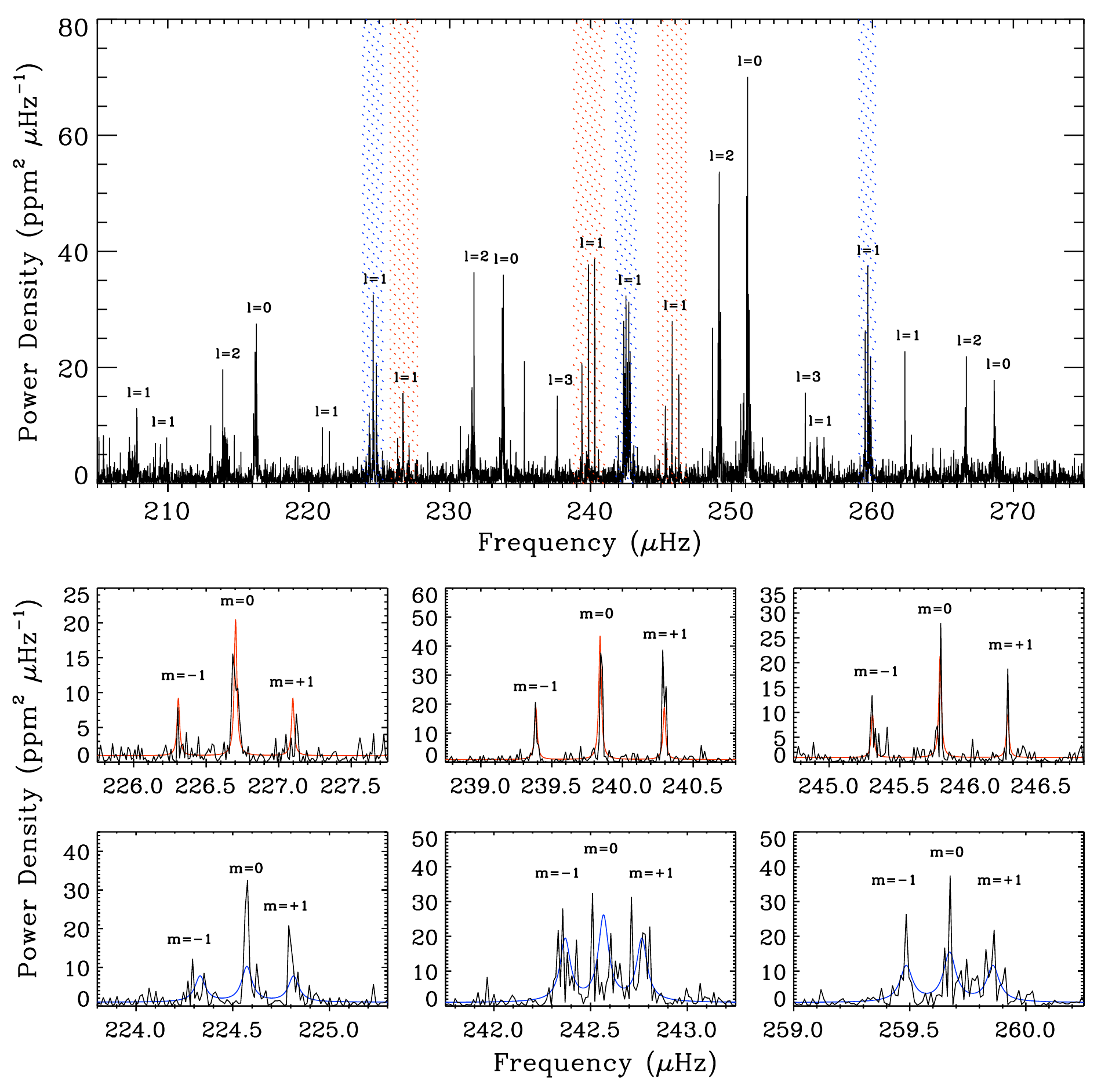}}
\caption{Top panel: Oscillation spectrum of Kepler-56. Gravity-dominated and
pressure-dominated mixed dipole modes are highlighted as red and blue, 
respectively. Bottom panels: Close-up of dipole modes highlighted in the 
top panel. Each modes is split into a triplet by rotation, demonstrating that the 
spin axis of Kepler-56 is misaligned with the orbital axis of the 
two transiting planets. From \citet{huber13b}.}
\label{seismo:fig6}
\end{center}
\end{figure}

Stellar inclinations are important for studying the architecture and 
dynamical history of transiting exoplanets by constraining the 
angle between the stellar spin axis and the planetary orbit axis 
(the obliquity). Since the presence of transits shows that the orbital axis is 
perpendicular to the line of sight, a low stellar inclination automatically implies a 
misalignment of orbital plane and the equatorial plane of the star 
(a high obliquity)\footnote{Conversely, an inclination near 90 degrees does not 
imply a low obliquity in the absence of a measurement of sky-projected stellar spin-orbit 
angle.}. High obliquities are frequently observed in stars hosting hot Jupiters 
\citep[e.g.][]{winn10}, while 
stars hosting multiple coplanar planets have been observed to have low 
obliquities \citep[e.g.][]{sanchis13}. 
This has been taken as evidence that the formation of hot Jupiters is related to 
dynamical interactions, rather than migrations through a protoplanetary disk.

High-precision photometry by \textit{Kepler} and CoRoT have enabled 
asteroseismic stellar inclination measurements for several exoplanet systems 
\citep{chaplin13c,gizon13,benomar14,vaneylen14,lund14}. An 
intriguing example is Kepler-56, a red giant hosting two transiting planets 
initially confirmed through transit-timing variations \citep{steffen12}. The Kepler-56 
power spectrum shows dipole modes which are split into triplets 
(Figure \ref{seismo:fig6}). Modeling the frequency triplets yielded an inclination 
of $47\pm6$ degrees, 
and demonstrated the first stellar spin-orbit misalignment in a multiplanet 
system \citep{huber13b}.

\begin{figure}
\begin{center}
\resizebox{\hsize}{!}{\includegraphics{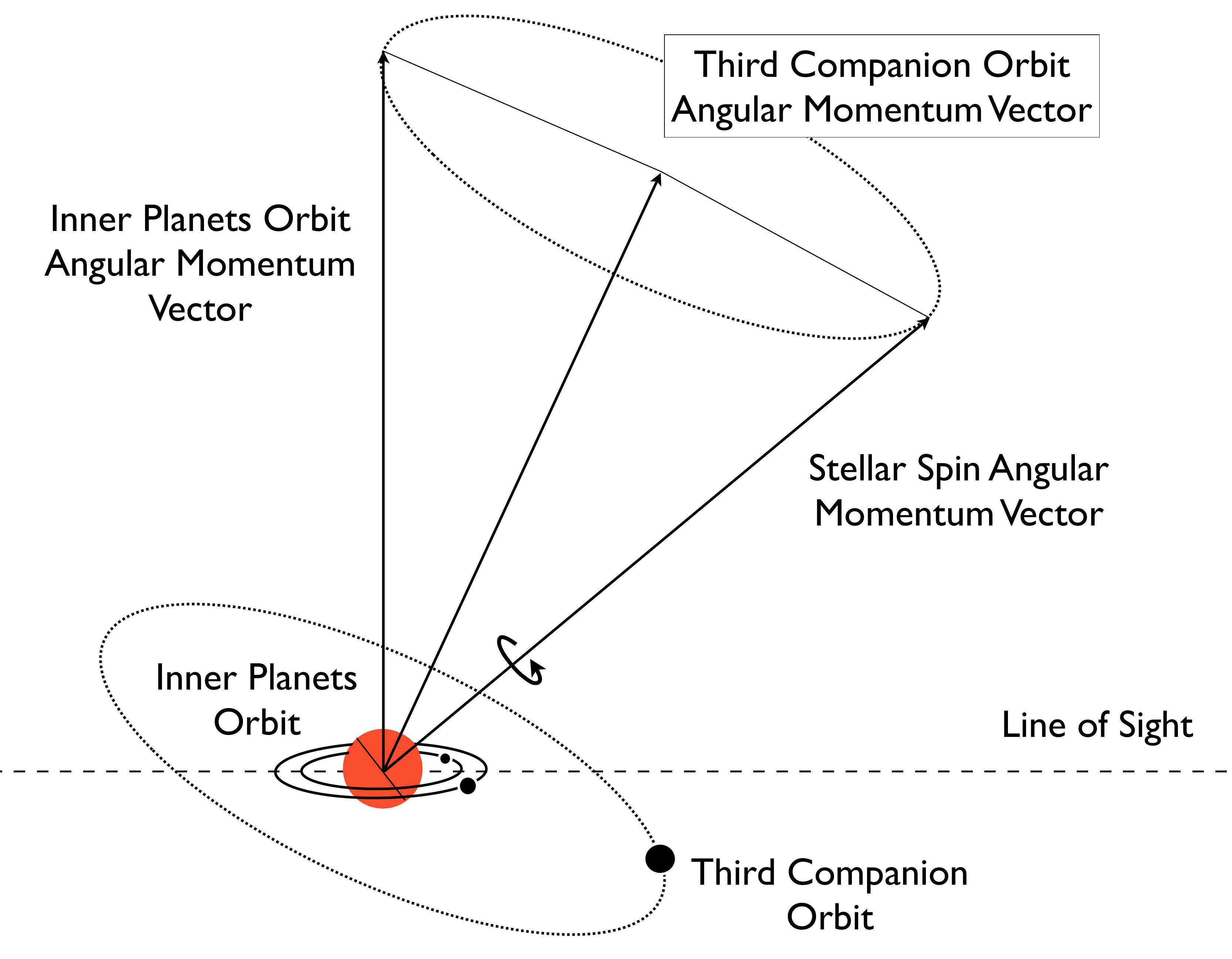}}
\caption{Graphical sketch of the Kepler-56 system. The torque of the outer companion 
causes a precession of the orbital axis of the inner transiting planets around the 
total angular momentum. The precession occurs 
at a different rate than the precession of spin axis of the host star, causing a 
periodic stellar spin-orbit misalignment. 
Sizes are not to scale. From \citet{huber13b}.}
\label{seismo:fig7}
\end{center}
\end{figure}

Follow-up radial velocity observations of Kepler-56 with Keck/HIRES revealed a long-term 
trend due to a massive companion on a wide orbit. Assuming a significant mutual inclination 
between the outer companion and the inner transiting planets, dynamical simulations 
showed that the misalignment can be explained by the precession of the orbital axis of the 
inner planets due to the torque of the wide companion (Figure \ref{seismo:fig7}), a 
scenario which has previously been proposed theoretically \citep{mardling10,kaib11,batygin12}. 
Further simulations confirmed that the misalignment in Kepler-56 is 
consistent with a dynamical origin due to an inclined perturber \citep{li14}. 

Kepler-56 demonstrated that spin-orbit misalignments are not confined to hot Jupiter 
systems, which has since also been suggested 
for the innermost planet of the 55 Cancri system based on Rossiter-McLaughlin observations 
\citep[][but see also L{\'o}pez-Morales et al., 2014]{bourrier14}. Further 
asteroseismic inclination measurements of exoplanet host stars, which are independent of 
the properties of planets and hence applicable to a wide range of systems, 
will reveal whether spin-orbit misalignments in multiplanet systems are common.
\nocite{lopezmorales14}

\section{Conclusions}

High-precision space-based photometry has triggered a revolution in 
asteroseismology of cool stars over the past few years. 
Highlights of recent asteroseismic 
discoveries presented in this review include the study of the internal rotation evolution 
of red giant stars and the precise characterization 
of the properties and architectures of exoplanet systems. Currently operating and 
planned space missions such as K2 \citep{howell14}, TESS \citep{ricker14} and 
PLATO \citep{rauer14}, as well as ground-based 
networks such as SONG \citep{grundahl08} and LCOGT \citep{brown13}, promise a bright future 
for asteroseismic studies of cool stars and their planets over the coming decades.

%
%


%
%




\acknowledgments{
Many thanks to Gerard van Belle and the LOC \& SOC for a fantastic week in Flagstaff. 
Financial support was provided by NASA grant NNX14AB92G issued 
through the Kepler Participating Scientist Program and the Australian Research Councils 
Discovery Projects funding scheme (project number DEI40101364).
}

\end{document}